# Scaling Studies for Efficient Parameter Search and Parallelism for Large Language Model Pre-training

MICHAEL BENINGTON*, LEO PHAN*, CHRIS PIERRE PAUL*, and EVAN SHOEMAKER*, Oak Ridge Institute For Science And Education, USA

PRIYANKA RANADE†, TORSTEIN COLLETT†, GRANT HODGSON PEREZ†, and CHRISTOPHER KRIEGER†, Laboratory for Physical Sciences, USA

## 1 ABSTRACT

AI accelerator processing capabilities and memory constraints largely dictate the scale in which machine learning workloads (e.g., training and inference) can be executed within a desirable time frame. Training a state of the art, transformer-based model today requires use of GPU-accelerated high performance computers with high-speed interconnects. As datasets and models continue to increase in size, computational requirements and memory demands for AI have increased proportionally [3]. A phenomenon observed among recent state of the art transformer-based Large Language Models (LLMs) is that the use of larger training corpora and models continue to improve accuracy for a diverse set of downstream tasks, encouraging the community to experiment in building larger models. For example, GPT-4 delivers significantly higher accuracy in downstream tasks such as question-answering and multi-task language understanding over its immediate predecessor, GPT-3 [4]. These advances have stimulated the growing demand for computing power very quickly, and as a result, the ML community is facing an impending slowdown in the ability to train larger foundation models. In 2018, OpenAI conducted an AI power and cost study inspired by Moore's law, and suggested that AI training runs have increased exponentially with a 3.4 month doubling time, compared to Moore's law which has a 2 year doubling period [3]. This doubling time suggests that training a 100 trillion parameter model would need 83,000 GPUs running for a year and would cost over $20 billion [1, 3].

These challenges have inspired the development of distributed algorithm and circuit-based optimization techniques that enable the ability to progressively scale models in multi-node environments, efficiently minimize neural network cost functions for faster convergence, and store more parameters into a set number of available resources. Specifically, the machine learning community is moving towards 3 major focus areas:

(1) Designing custom computer architectures.
(2) Parallel/Distributed machine learning algorithm development.
(3) Developing novel model architecture modifications for efficient learning.

In our research project, we focus on (2) *Parallel/Distributed machine learning algorithm development*, specifically for optimizing the data processing and pre-training of a set of 5 encoder-decoder LLMs, ranging from 580 million parameters to 13 billion parameters across an 8 node 8-A100 DGX system. We performed a fine-grained study to quantify the relationships between three ML parallelism

---

*All authors contributed equally to this research.
†Research Mentors

Authors' addresses: Michael Benington, mrbenin@lps.umd.edu; Leo Phan, qmphan@lps.umd.edu; Chris Pierre Paul, cppaul@lps.umd.edu; Evan Shoemaker, ecshoem@lps.umd.edu, Oak Ridge Institute For Science And Education, Oak Ridge, TN, USA; Priyanka Ranade, psranad@lps.umd.edu; Torstein Collett, tccolle@lps.umd.edu; Grant Hodgson Perez, gmhodgs@lps.umd.edu; Christopher Krieger, christopher.d.krieger.civ@mail.mil, Laboratory for Physical Sciences, College Park, MD, USA.

methods: *data, model, and tensor* parallelism (specifically exploring Microsoft DeepSpeed Zero Redundancy Optimizer (ZeRO) stages) [7] and 30 hyperparameter optimization techniques (finding the effective batch size, scaling learning rate, selecting an efficient optimizer, etc).

Previous work has mainly explored benefits of ML parallelism and hyperparameter optimization techniques independently or for a defined resource allocation space (#nodes or accelerators). Our project consists of observing hyperparameter and parallelism relationships in a scaling environment (continuously adding nodes and accelerators to the search space). The hyperparameter search space initially consisted of 30 different hyperparameter dimensions. Each *run* represents a single model configuration with one, or a selected *subset* of the total hyperparameters. The selection process was determined through a *prune and combine* procedure described in the next paragraph. Runs were executed on single accelerators, single-node, and multi-node environments. We observed changes in two main performance metrics: (1) Seconds per step, which we use to project an expected time-to-train and (2) Changes in model loss and accuracy to predict steps required for convergence.

Following these evaluation metrics, our study implemented a funneled hyperparameter search approach, in which we first broadly observed changes to single parameters at a time, while keeping all others constant on a single node. For every parameter that was changed, or added, a new template was created. We then pruned certain parameters and combined the best resulting templates across the first phase and created combination templates, containing a mixture of ML parallelism and optimal, observed hyperparameters. We continued this prune and combine process until we found a set of hyperparameters that resulted in the best performance for a given range of models to test in multi-node environments. We selected a total of 15 templates to benchmark across 4-8 node tests.

We hypothesized that using varying forms of parallelism (e.g., progressing through the DeepSpeed ZeRO stages), and adding additional nodes would increase the number of steps per second (e.g., leading to decreased time to train to converge to an expected accuracy). However, we confirmed that using the highest form of DeepSpeed (stage 3) does not improve training speed, especially when scaling to additional resources, because it comes with increased communication overhead. For example, DeepSpeed enables model parallelism, which allows us to fit more parameters given a set number of resources, but simultaneously has been shown to increase inter-node communication during ZeRO stage progression [6]. For example, Table 1, shows time to train results for DeepSpeed ZeRO stage 2 (partitioning the reduced 32-bit gradients for updating the model weights) and 3 (partitioning the 16-bit model parameters) while sclaing across 2, 4, and 8 nodes for mt5-XXL (13B paramter) pretraining. We use the effective batch size, linear learning rate, and keep step count constant for all tests, to ensure direct comparison for each test case [5]. We report the fastest seconds per step observed for each test case. For each node count, stage 2 resulted in lower seconds per step in comparison to stage 3. Additionally, model training across 8 nodes was slower than both 2 and 4 nodes, while 4 nodes was faster than both. In future work, we hope to perform an inter-node communication study to assess potential bottlenecks while parallelizing across the entire, 8 node system in comparison to 4 nodes.

This is one example study where we found that scaling the number of nodes during DeepSpeed training of the mt5 XXL model does not correlate to the speedup in training time. The addition of extra nodes failed to provide a speedup in training. One of the reasons we believe the addition of extra nodes contrasted the scale to a speedup in training was due to the increased communication overhead between nodes. This stresses the importance of having sufficient interconnect between nodes to address the communication overhead, for example, to allow for DeepSpeed's 1) all-gathers for collection, 2) scatter for partioning, and 3) CPU offloading [2]. Furthermore, the lack of parallelism in dataloaders that provide the training data to each node may cause slow down in training speed when scaling to multiple nodes.

Scaling Studies for Efficient Parameter Search and Parallelism for Large Language Model Pre-training    3|  | **Number of Nodes** | | |
|---|---|---|---|
| **DeepSpeed Stage** | 2 | 4 | 8 |
| 2 | 20.38 | 12.00 | 31.42 |
| 3 | 25.78 | 23.25 | 38.86 |

Table 1. This table shows the training seconds per step for different combinations of DeepSpeed stages and number of nodes.

We also found that based on 205 trials, there are too many hyperparameters and training paradigms to have a "one fits all" recipe. Certain hyperparameter combinations can work well in certain scenarios, but in others be ineffective. A key implication of this work is that training a custom model or even fine-tuning an existing one, continues to be a resource and time-intensive task that demands much trial and error. In future work, we hope to translate our findings into a novel hyperparameter search algorithm specifically made for scaling environments.